\documentclass[12pt]{spieman}  
\usepackage{amsmath,amsfonts,amssymb}
\usepackage{graphicx}
\usepackage{setspace}
\usepackage{tocloft}
\usepackage{lineno}
\usepackage{hyperref}
\usepackage{xcolor}

\title{Commissioning of the large-scale lead tungstate scintillating calorimeter}

\author[a,*]{Alexander Somov}
\author[a]{Vladimir V. Berdnikov}
\author[b]{Liping Gan}
\author[b]{Puthiya V. Laveen}
\author[a]{Andrew Smith}
\author[a]{Simon Taylor}
\author[c]{Hakob Voskanyan}

\affil[a]{Thomas Jefferson National Accelerator Facility, Newport News, Virginia 23606, USA}
\affil[b]{University of North Carolina at Wilmington, Wilmington, NC 28403, USA}
\affil[c]{A. I. Alikhanian National Science Laboratory (Yerevan Physics Institute), 0036 Yerevan, Armenia}

\cftpagenumbersoff{figure}
\cftpagenumbersoff{table} 
\begin{document} 
\maketitle

\begin{abstract}
This article reports on the installation and initial commissioning of a large-scale lead tungstate (PbWO$_4$) scintillating crystal calorimeter developed for high-rate photon detection and precise energy measurement. The calorimeter comprises 1,596 high-granularity, high-resolution scintillating crystals optimized for electromagnetic-shower detection over a wide energy range. Scintillation light from each crystal is read out by Hamamatsu R4125 photomultiplier tubes equipped with a custom voltage divider and front-end amplifier to ensure stable gain at high rates. All calorimeter modules were fabricated and characterized using an LED-based optical test system prior to installation to verify uniformity and photodetector performance. After installation, the ECAL was fully integrated into the experiment data acquisition and energy-based trigger systems. The optical response of the modules was equalized using the light-monitoring system, cosmic-ray muons, and photons from Compton-scattering events. Commissioning results demonstrate a reliably calibrated optical response and stable detector performance during the first run. These results validate the calorimeter design and commissioning methodology for large-scale scintillator-based photonic instrumentation.

\end{abstract}


{\noindent \footnotesize\textbf{*}Address all correspondence to Alexander Somov,  \linkable{somov@jlab.org} \\ 
P.V. Laveen: Present address: Center for Applied Isotope Studies, University of Georgia, Athens, GA 30602, USA }

\begin{spacing}{1} 

\section{Introduction}
\label{sec:intro}  

Electromagnetic calorimeters based on lead tungstate (PbWO$_4$) crystals are widely used in experimental nuclear and high-energy physics for the reconstruction of electromagnetic showers produced by photons and electrons. Owing to the unique properties of PbWO$_4$~\cite{pwo_prop,pwo_prop1,pwo_prop2,pwo_prop3}, these calorimeters provide an excellent balance of fast response, compactness, high precision, and radiation tolerance, making them well suited for high-rate environments. The first large-scale implementation of PbWO$_4$ calorimetry was realized in the CMS~\cite{cms} and ALICE~\cite{alice} experiments at CERN. This technology was later adopted in several calorimeter systems at Jefferson Lab~\cite{hycal,hps,clas12,nps}. PbWO$_4$ crystals were also selected for the  instrumentation of the PANDA experiment~\cite{panda} at GSI Helmholtz Center for Heavy Ion Research, as well as for a calorimeter to be installed at the future  electron–ion collider facility~\cite{eic} at Brookhaven National Laboratory.

The PbWO$_4$ electromagnetic calorimeter (ECAL)  was chosen for the Jefferson Lab Eta Factory (JEF) experiment~\cite{jef,somov_jef} and subsequently developed and installed as part of the GlueX detector~\cite{gluex}. The calorimeter measures electromagnetic showers from multi-photon final states over a wide energy range from 100~MeV to 10~GeV, enhancing the GlueX photon detection capabilities and supporting high-rate operation with precise energy measurement.

The ECAL comprises 1,596 lead tungstate scintillating crystals that replaces the inner region of the existing GlueX forward lead-glass calorimeter (FCAL)~\cite{gluex}. Lead tungstate offers several advantages over lead glass, including a smaller radiation length and Moli$\grave{\rm e}$re radius, higher light yield, and superior radiation hardness. These characteristics result in approximately a factor of two improvement in energy resolution ($\Delta E / E \sim 3\%/\sqrt(E)$ at small energies) and provide a fourfold increase in detector granularity compared to the previous detector. Scintillation light from each crystal is detected by a Hamamatsu R4125 photomultiplier tube (PMT), each equipped with a custom voltage divider and front-end amplifier to ensure linear gain and stable operation at rates up to 500~kHz per module. The PMT electronics were optimized using a detector prototype developed and tested at Jefferson Lab~\cite{ccal_nim}.

ECAL module construction began in 2022 under a comprehensive quality assurance program for PbWO$_4$ crystals procured from two vendors: the Shanghai Institute of Ceramics (SICCAS) and CRYTUR in the Czech Republic~\cite{crystals}. Each assembled detector module, including the PMT base, was optically characterized using an LED-based light source prior to installation in the GlueX detector.

A dedicated monitoring system (LMS) was implemented to test installed modules and  monitor detector stability during operation. The system continuously monitor changes in the optical response arising from temperature-dependent crystal light yield, variations in PMT gains, and radiation-induced crystal degradation.

Following installation, the ECAL  was integrated into the GlueX data acquisition and energy-based trigger system, operating at $\sim\!70$~kHz. 
Uniformity of the detector optical response is essential for reliable trigger performance and requires module-by-module equalization at the 5-10$\%$  level via adjustment of the PMT high voltages. 

\begin{figure} [t]
\begin{center}
\begin{tabular}{c} 
\includegraphics[width=16.0cm]{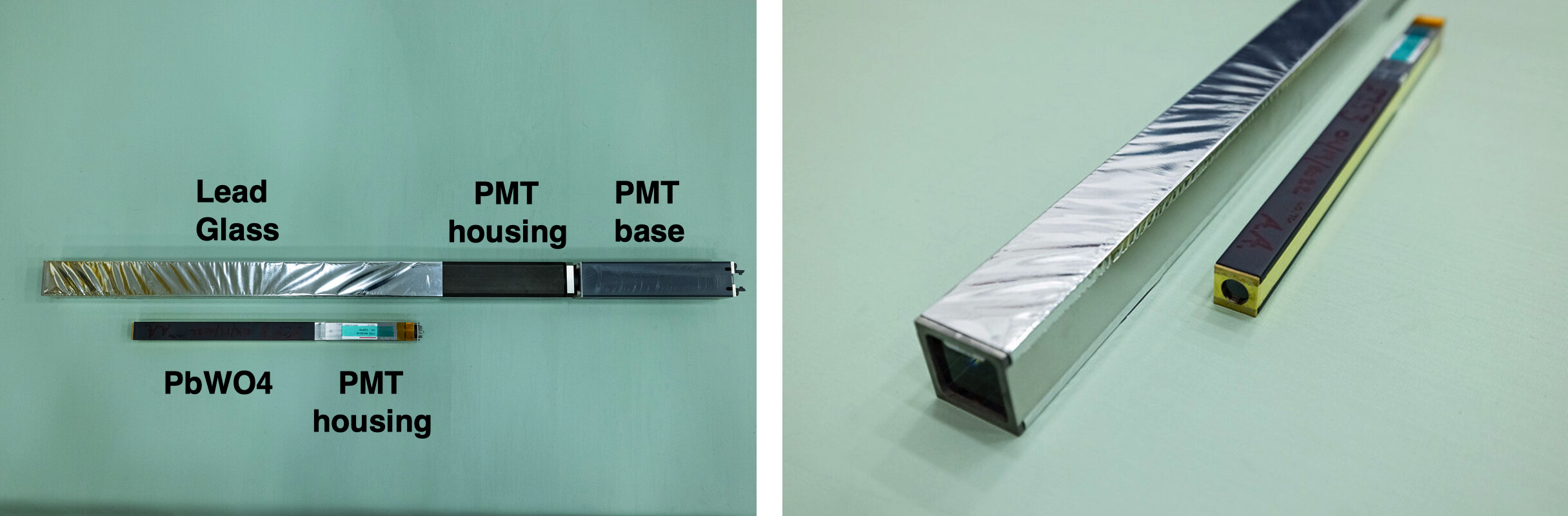}
\end{tabular}
\end{center}
\caption[example] 
{ \label{fig:ecal_module} 
Lead glass (long)  and lead tungstate (short) calorimeter modules.}
\end{figure} 

During detector commissioning, the equalization of the optical and electronic response was performed in several stages: (1) the light monitoring system was used to measure the dependence of signal amplitude on PMT high voltage, which is used to derive voltage corrections; (2) scintillation signals from minimum-ionizing cosmic-ray muons were used to establish initial PMT high-voltage settings; and (3) electromagnetic showers from photons with well-defined energies produced via Compton scattering were reconstructed and used to refine PMT high voltages and improve detector response uniformity. The preliminary commissioning results were presented in our earlier SPIE Proceedings paper~\cite{ecal_install}. Subsequent offline calibration using physics data further improved the overall energy resolution. Throughout commissioning, the detector demonstrated stable and reliable performance.

This article focuses on the optical and instrumentation aspects of the ECAL, including module design, light monitoring system, and commissioning results, demonstrating a robust methodology  for deploying and maintaining large-scale scintillator-based photonic detection systems. The article is organized as follows: Section~\ref{sec:ecal_modules} describes the design, assembly and testing of the ECAL modules; Section~\ref{sec:installation} presents the detector installation and its integration into the GlueX infrastructure; and Section~\ref{sec:ecal_commission} discusses the initial commissioning results of the new detector.

\section{ECAL module assembling and testing}
\label{sec:ecal_modules}

The calorimeter features a modular design~\cite{somov_ecal,ccal_nim}. A total of 1596 modules form a $40\times40$ array, with a central of $2\times2$ hole for the photon beam. Each lead tungstate crystal, measuring 2.05 cm × 2.05 cm × 20 cm, is wrapped in Enhanced Specular Reflector foil (produced by ${\rm 3M}^{\rm TM}$) and light-tight Tedlar. Scintillation light is collected using a Hamamatsu R4125 photomultiplier tube, which is wrapped in mu-metal foil and placed inside an Amumetal cylinder and soft iron housing to reduce the effects of stray magnetic fields from the GlueX solenoid. The PMT is coupled to the crystal via an acrylic light guide, which extends the magnetic shielding beyond the photocathode face of the PMT. One end of the light guide is glued to the PMT, while the other is attached to the crystal using optical silicone rubber. An ECAL module is shown in Fig.~\ref{fig:ecal_module}, together with a lead glass module. The lead-glass module has a similar overall design: a rectangular lead glass block, 4 cm × 4 cm × 45 cm, in which Cherenkov light is produced by electromagnetic showers, is coupled to an FEU-84 PMT. The significantly shorter radiation length of PbWO$_4$ allows the ECAL to achieve a more compact detector design and improved granularity.

The ECAL project commenced with the characterization and quality control of lead tungstate crystals. This was followed by preparation of components required for module assembly, including gluing photomultiplier tubes  to light guides, shaping reflective foils, and producing silicone optical pads ("cookies"). A total of 1620 detector modules, including spares, were assembled at Jefferson Lab.

The original voltage divider for the R4125 photomultiplier tube, which supplies voltages to the dynodes and provides signal readout, was modified at Jefferson Lab by adding two transistors to the divider chain and an amplifier with a gain of approximately three~\cite{divider}. The amplifier was mounted directly on the divider's printed circuit board. These modifications made it possible to reduce the operating high voltage and, consequently, the anode current, thereby improving performance under high-rate conditions. The amplifier requires an additional $\pm 5{\rm V}$ power supply. The modified divider is referred to as the PMT active base.  Signal pulses from the PMTs are digitized using an analog-to-digital converter (ADC) operating at a sampling rate of 250 MHz.

\begin{figure} [b]
\begin{center}
\begin{tabular}{c} 
  \includegraphics[width=12.0cm]{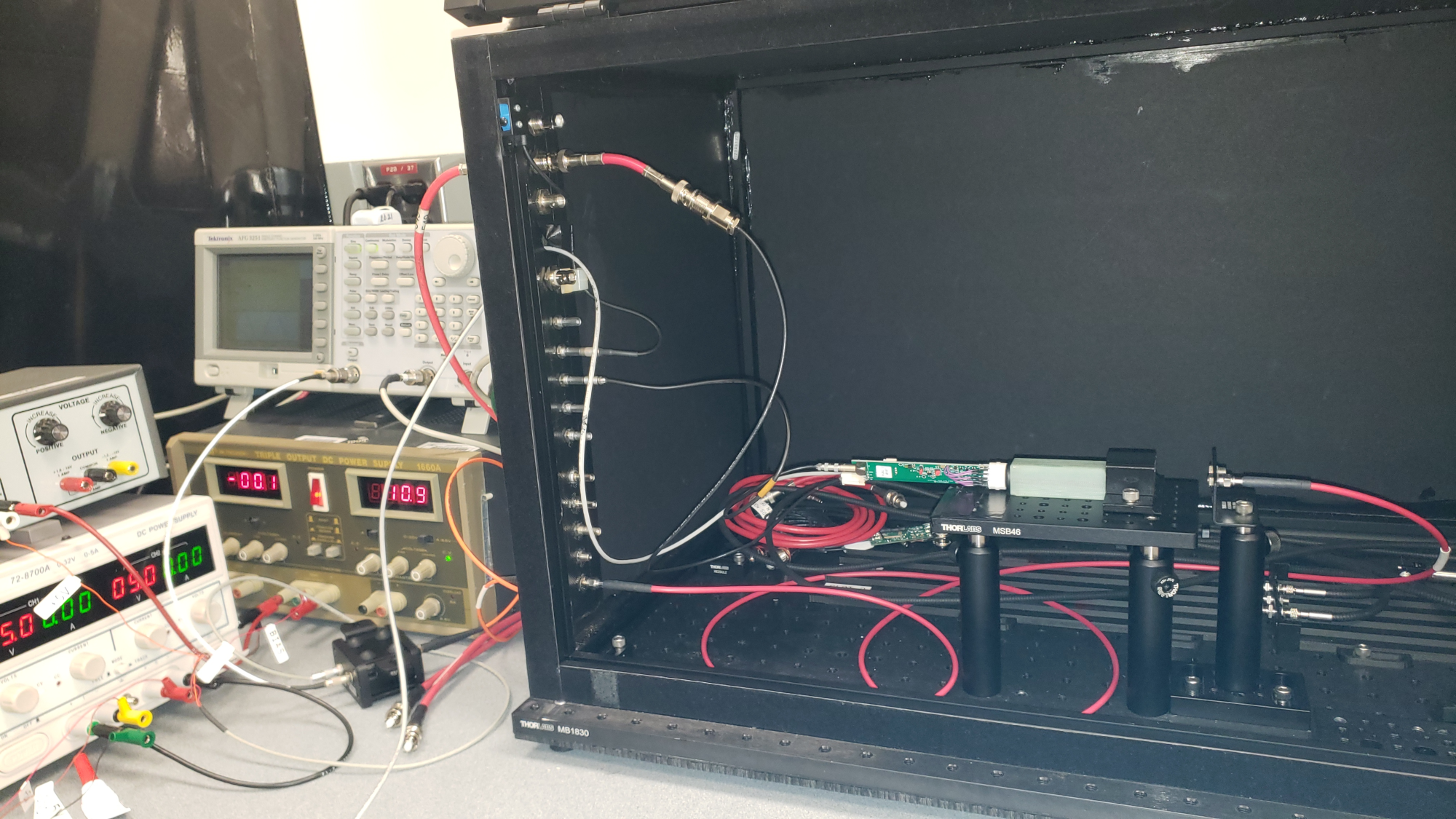}
\end{tabular}
\end{center}
\caption[example] 
{ \label{fig:testing_dividers} 
Setup for testing ECAL PMT active bases and assembled modules using the LED-based light-monitoring system. }
\end{figure} 
\begin{figure} [htb]
\begin{center}
\begin{tabular}{c} 
  \includegraphics[width=12.0cm]{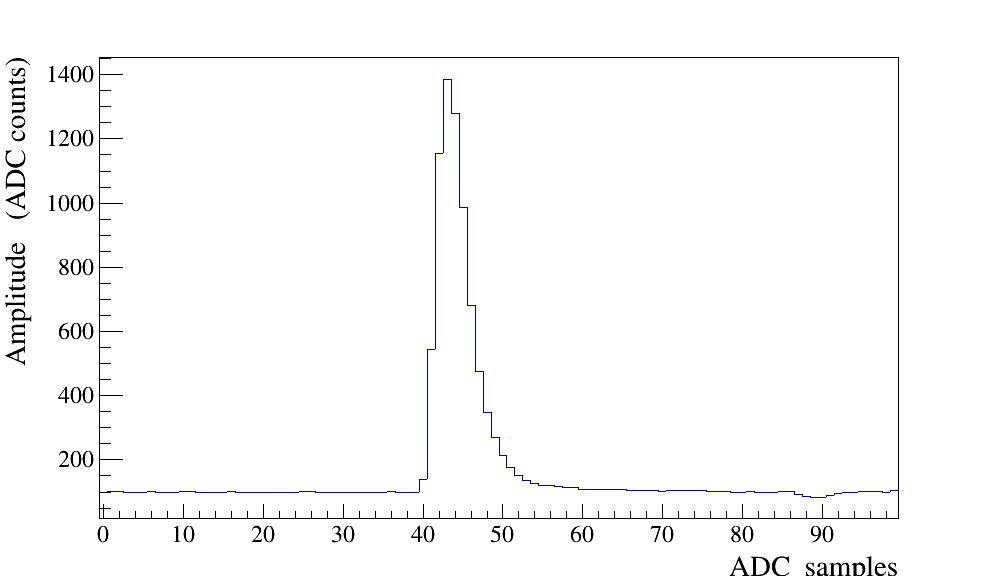}
\end{tabular}
\end{center}
\caption[example] 
{ \label{fig:signal_waveform} 
Signal waveform digitized by a flash ADC induced by the LED-based light-monitoring system. The flash ADC operates at a sampling rate of 250~MHz, corresponding to a 4~ns sample interval.
}
\end{figure} 
All assembled ECAL modules were tested using a pulser system based on a blue light-emitting diode (LED). Each module was placed inside a light-tight enclosure, and light from the LED was delivered to the front face of the module via an optical fiber. The test setup is shown in Fig.~\ref{fig:testing_dividers}. Signal waveforms generated  by the LED were recorded using an oscilloscope. The LED-based light-monitoring system was subsequently used to verify the performance of all modules and readout electronics installed in the detector. An example of the digitized signal waveform is shown in Fig.~\ref{fig:signal_waveform}. The same setup was also used to test the PMT active bases. Each base was connected to a reference PMT operated at a fixed high voltage of 1~kV. Both the pulse amplitude and the current drawn by the active base were measured, with the latter being approximately 400~\textmu A.

\section{Detector Installation}
\label{sec:installation}

\begin{figure} [htb]
\begin{center}
\begin{tabular}{c} 
  \includegraphics[width=14cm]{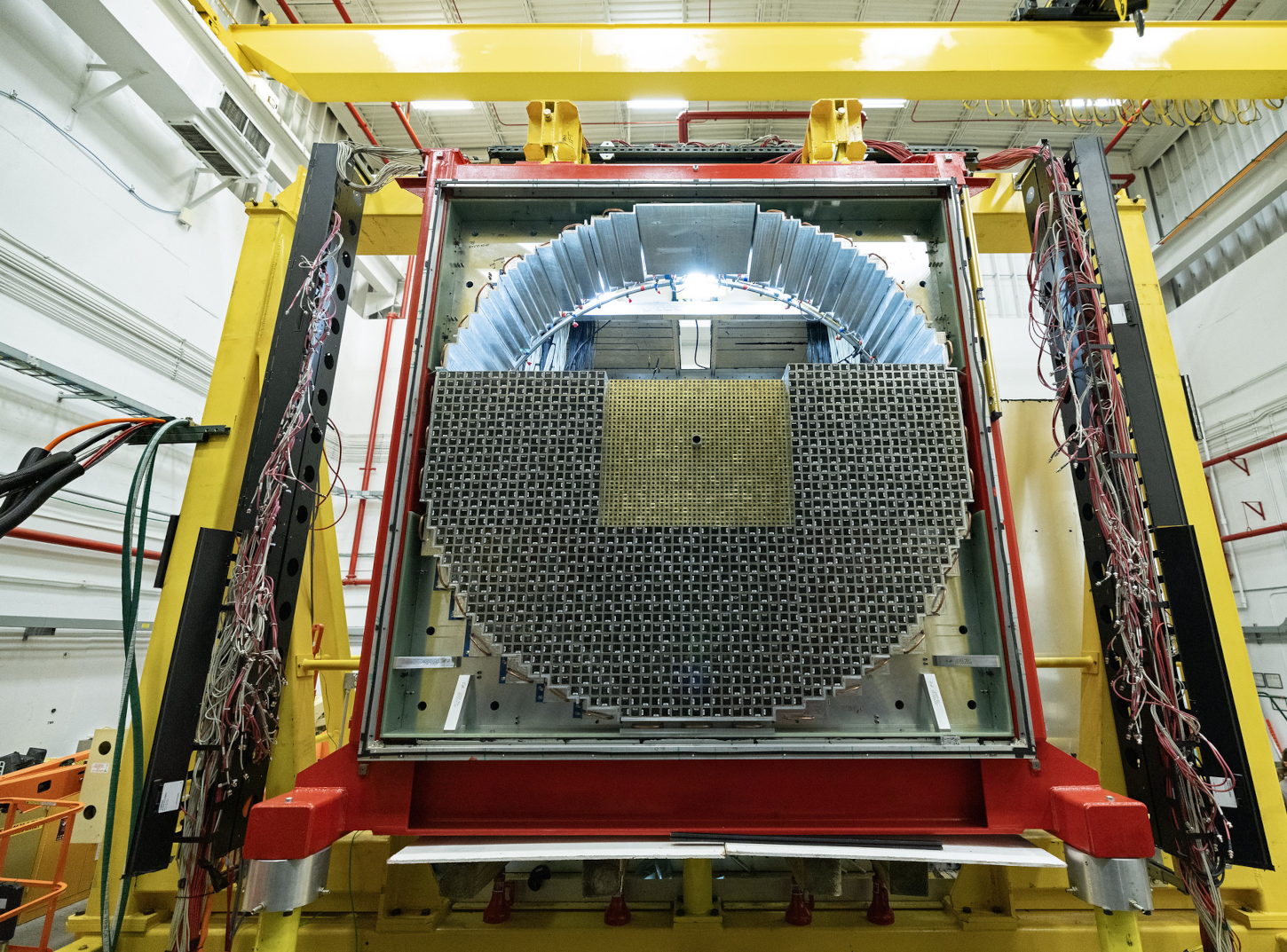}
\end{tabular}
\end{center}
\caption[example] 
{ \label{fig:ecal_installation} 
Installation of the ECAL in experimental Hall~D. The ECAL modules at the center are surrounded by lead-glass modules.}
\end{figure} 
\begin{figure} [htb]
\centering
\begin{tabular}{c} 
\includegraphics[width=16cm]{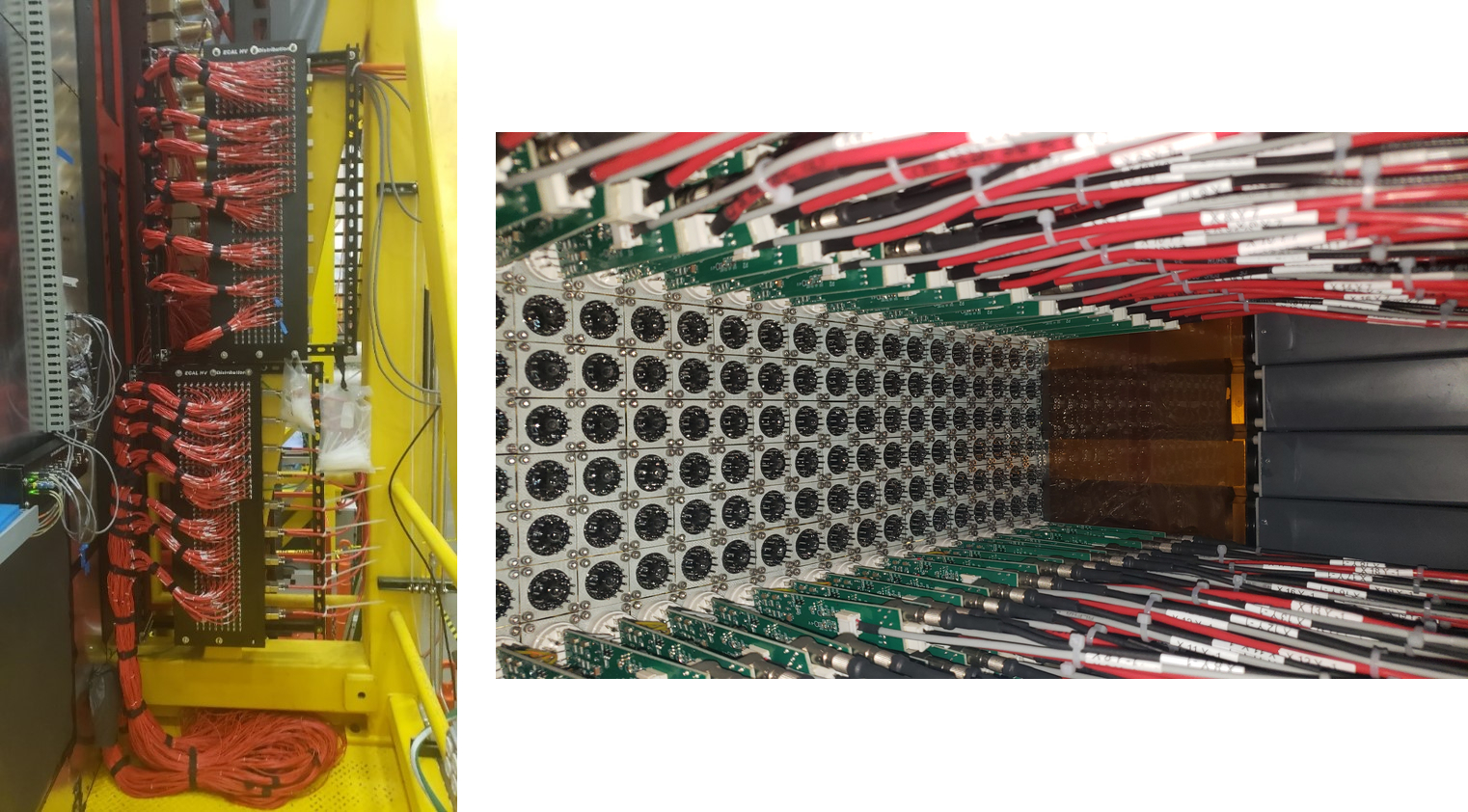}
\end{tabular}
\centering
\caption[example] 
{ \label{fig:ecal_divider} 
Two high-voltage distribution boards installed near the detector housing (left). Connection of active bases to the PMT sockets (right). Each base is connected to three cables: signal, high voltage, and low voltage.}
\end{figure}  

The installation of the ECAL in the experimental hall began in April 2023 with the removal of 2,800 lead glass-modules from the detector frame. A new support structure, consisting of aluminum blocks with embedded water-cooling pipes, was installed. The cooling system is supplied by two chillers, maintaining a constant temperature, which is critical for detector performance because the light yield of PbWO$_4$ crystals is temperature dependent. The lead-glass modules located beneath the ECAL were installed first, followed by the stacking of the ECAL modules. The installation process is shown in Fig.~\ref{fig:ecal_installation}. The detector was subsequently enclosed in a light-tight, thermally insulated housing to maintain a stable operating temperature.

The installation of the ECAL cables was performed in multiple steps. Signal, high-voltage (HV) and low-voltage (LV) cables, each approximately 50~ft long, were routed from the electronics modules to distribution boards mounted on the frame of the detector housing. The distribution boards provided an organized interface between the electronics and the detector, allowing long external cables to be connected outside the detector housing and short internal cables to attach to the PMT active bases, which simplifies installation and maintenance. Outside the housing, 1596 signal cables were connected to one hundred 16-channel ADCs modules positioned in seven VXS crates. Thirty-six multiwire HV cables were connected to thirty-six 48-channel CAEN 7030N modules installed in three CAEN HV SY4527 mainframes. Low voltage was supplied by two MPOD modules positioned in a Wiener crate and distributed to the distribution boards via four cables. In total, approximately 6,500 cables were installed for the ECAL detector. Two high-voltage distribution boards with connected cables are shown in the left plot of Fig.~\ref{fig:ecal_divider}. Inside the detector housing, short cables (up to 16~ft) were used to connect the PMT active bases to the distribution boards. Each active base received three individual cables: signal, HV, and LV. In the final stage of installation, the cables were connected to the PMT active bases, which were then attached to the PMT sockets. The active base installation procedure is shown in the right plot of Fig.~\ref{fig:ecal_divider}. All cables were securely fastened to the vertical support chains to provide strain relief. To ensure proper PMT connections and cabling, voltages were applied after installing the dividers on each detector layer, and the module response was checked using the light-monitoring system, described in Section~\ref{sec_lms}. These tests were essential, as the high cable density limited access to the dividers after installation.

\section{ECAL commissioning}
\label{sec:ecal_commission}
\subsection{Integration into the data acquisition and trigger}

Commissioning of the ECAL required full integration of the detector into the GlueX trigger and data acquisition (DAQ) systems. The trigger is based on energy depositions in the ECAL, forward, and barrel (BCAL) electromagnetic calorimeters~\cite{gluex}. Since the ECAL replaces the inner region of the FCAL, its signals are combined with those from surrounding FCAL modules. The trigger condition can be expressed as
\begin{equation}
1.4\cdot (E_{\rm ECAL} + 0.5 \cdot E_{\rm FCAL}) + E_{\rm BCAL} > 1\;{\rm GeV},
\label{eq:trig}
\end{equation}
where $E_{\rm ECAL}$, $E_{\rm FCAL}$, and $E_{\rm BCAL}$ are the energies measured in the respective calorimeters. The coefficients in Eq.~\ref{eq:trig} determine the relative contribution of each calorimeter to the trigger energy and are optimized based on the physics requirements of the experiment. During high-rate operation, modules near the beam pipe are excluded to reduce low-angle background, while during calibration runs, all modules contribute.

The trigger is implemented on custom electronics designed at Jefferson Lab. Trigger processing begins in flash ADCs located in VXS crates compliant with the ANSI/VITA 41.0 standard. Signals from ECAL photomultiplier tubes are digitized, and the resulting waveforms are processed on a Field-Programmable Gate Array (FPGA) chip to extract pulse parameters such as amplitude, integral, and timing. This information is stored in a pipeline for subsequent readout. In parallel, the digitized waveforms from each flash ADC are transmitted to the VXS Trigger Processor (VTP) module located in the middle of each VXS crate. This VTP sums the digitized amplitudes (or energies) from all ADC modules within the crate and forwards the summed energy an via optical link to the Sub-System Processor (SSP) module in the trigger crate. The SSP then sums the energies from all ECAL VXS crates and transmits the total energy to the Global Trigger Processor (GTP), where Eq.~\ref{eq:trig} is evaluated to form the final trigger decision. Trigger signals are distributed to each VXS readout crate by the Trigger Supervisor (TS) module to initiate data readout. The TS can also directly accept triggers that do not require processing by the GTP, including triggers from the light-monitoring system, cosmic-ray triggers (described in Section~\ref{sec:calib_cosmic}), and other ancillary GlueX triggers.

\begin{figure} [htb]
\begin{center}
\begin{tabular}{c} 
\includegraphics[width=10cm]{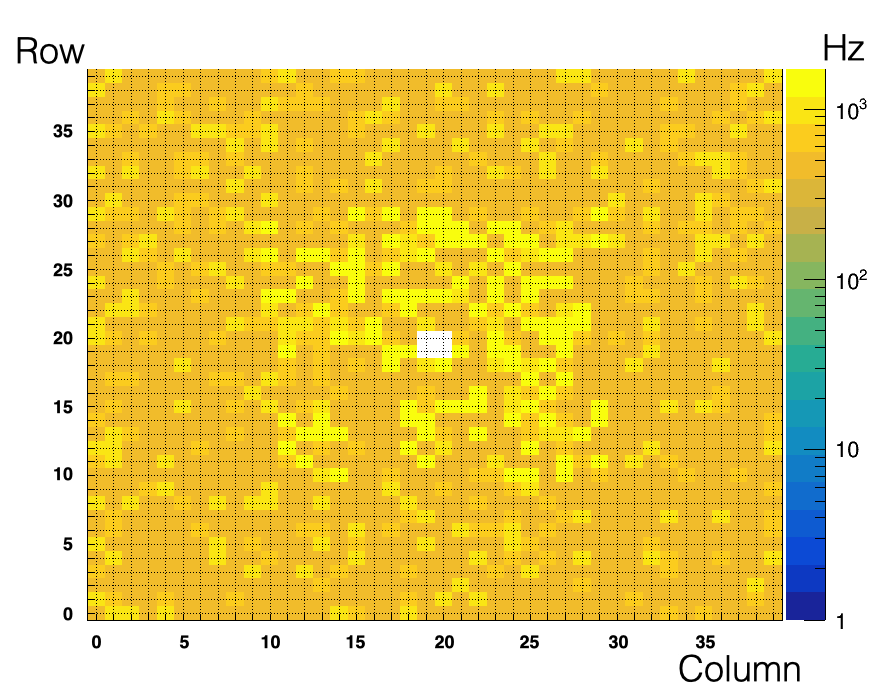}
\end{tabular}
\end{center}
\caption[example] 
{ \label{fig:lms_scalers} 
Signal pulse rate in the ECAL modules induced by the light monitoring system.}
\end{figure} 
\begin{figure} [b]
\begin{center}
\begin{tabular}{c} 
\includegraphics[width=10cm]{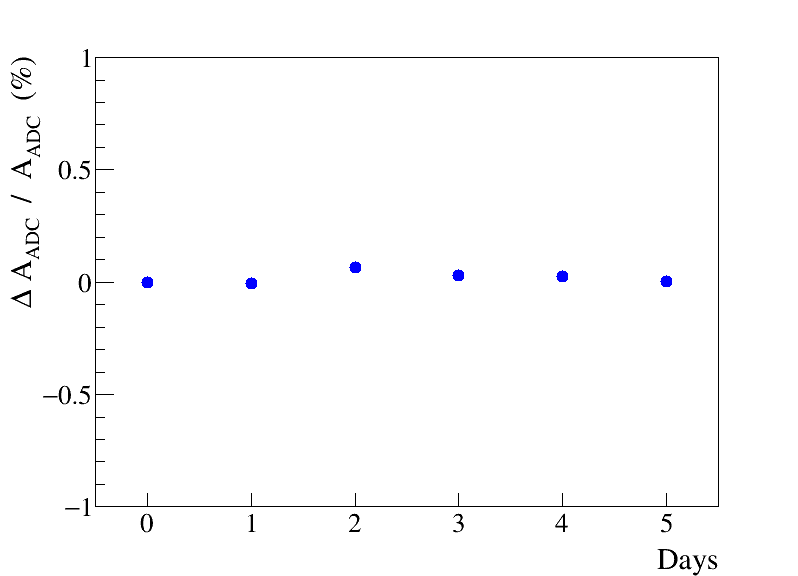}
\end{tabular}
\end{center}
\caption[example] 
{ \label{fig:lms_stab} 
Typical relative change of the ADC signal pulse amplitude in an ECAL module induced by the LMS over time.}
\end{figure} 
\begin{figure} [htb]
\begin{center}
\begin{tabular}{c} 
\includegraphics[width=12cm]{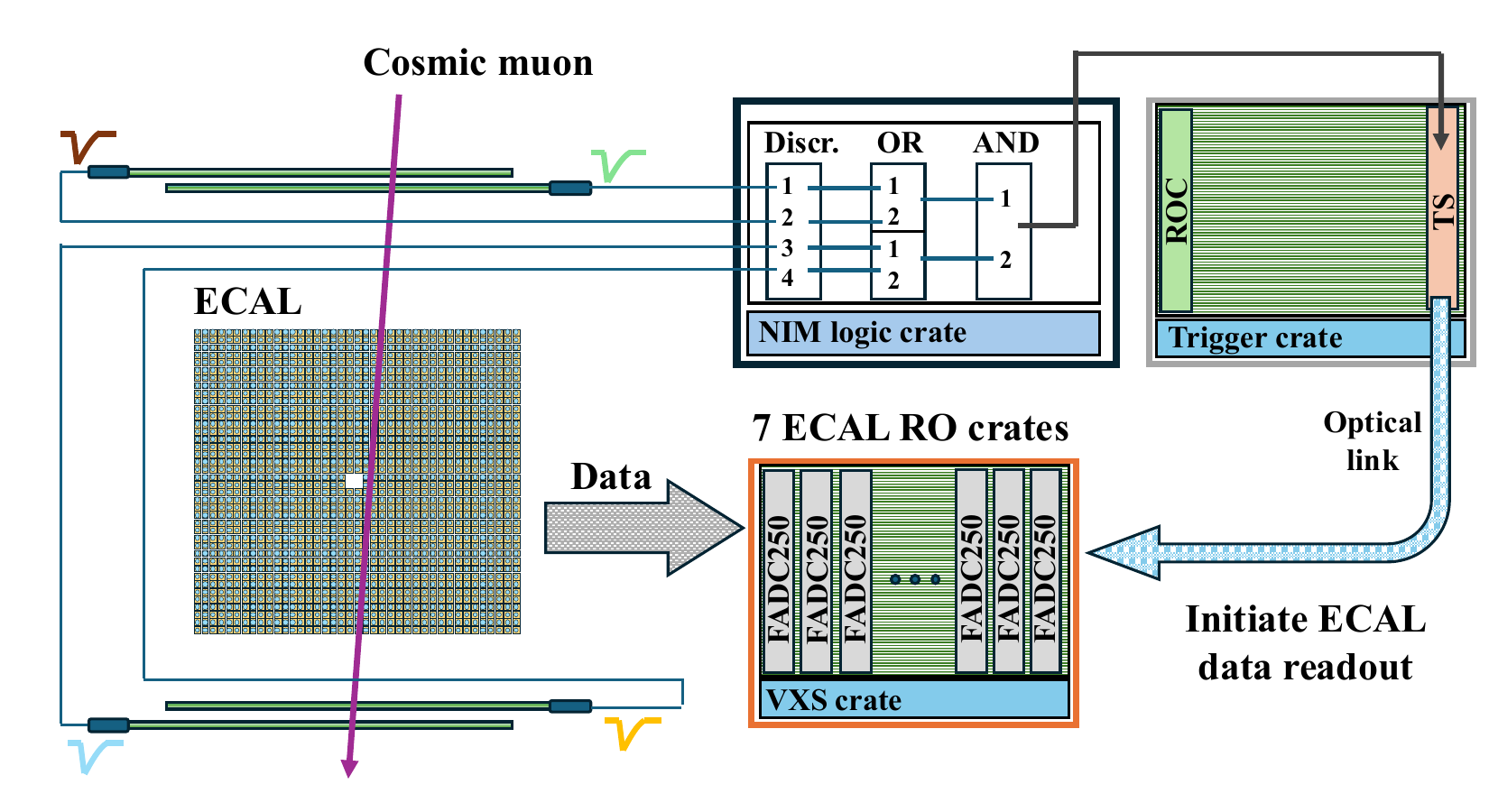}
\end{tabular}
\end{center}
\caption[example] 
{ \label{fig:cosmic_setup} 
Schematic view of the setup used to study cosmic rays.}
\end{figure} 

The trigger calculation assumes that all calorimeter modules are properly calibrated and their optical responses are equalized, such that a photon of a given energy produces the same flash-ADC amplitude in every module. Accurate detector calibration is therefore required and is described in the following section.

\subsection{Verification using light monitoring system}
\label{sec_lms}

The ECAL light monitoring system (LMS) was designed to monitor the detector's performance during operation.  The LMS generates light flashes that are used to monitor the detector's health and stability of the signal response over time. The light source consists of 25 blue LEDs, positioned inside an integrating sphere to ensure uniform light mixing. The mixed light is delivered to the face of each ECAL module via  500~\textmu m-diameter optical fibers. Each fiber is glued to the face of the crystal using UV-curable adhesive. LMS operating parameters, including flash rate and light amplitude, can be controlled remotely through the Experimental Physics and Industrial Control System~\cite{epics}. A full description of the ECAL LMS system is provided in Ref.~\cite{ecal_lms}.

During installation of the ECAL front-end electronics and cabling, the LMS was used to identify and resolve issues related to power supplies, cabling, and electronics. An algorithm implemented in the ADC FPGA enabled monitoring of the rate of LMS-induced signal pulses without operating the detector DAQ. This capability provided real-time feedback on the detector performance and significantly facilitated initial commissioning. An example of the LMS-induced signal pulse rates across the ECAL modules is shown in Fig.~\ref{fig:lms_scalers}, demonstrating that all ECAL modules were operational.

The LMS was integrated into the GlueX trigger system, allowing data to be recorded  using a dedicated trigger type. It was also used to study the dependence of the signal pulse amplitude on the applied high voltage for each module. These measurements were subsequently used to adjust the operating voltage of each individual module. 

The stability of the optical response of ECAL modules was evaluated during detector commissioning and data taking at nominal operating conditions using the LMS system. Fig.~\ref{fig:lms_stab}  shows an example of the typical relative change of the ADC amplitude induced by the LMS over time, defined as (A$_{\rm ADC}$ (t) -- A$_{\rm ADC}$ (0))/A$_{\rm ADC}$ (0), where A$_{\rm ADC}$ (0) and A$_{\rm ADC}$ (t) are the initial amplitude and the amplitude at time $t$, respectively. The observed relative change remains within $\pm 1\%$, demonstrating good stability of the ECAL response over approximately five days of data taking.

\begin{figure} [ht]
\begin{center}
\begin{tabular}{c} 
\includegraphics[width=10cm]{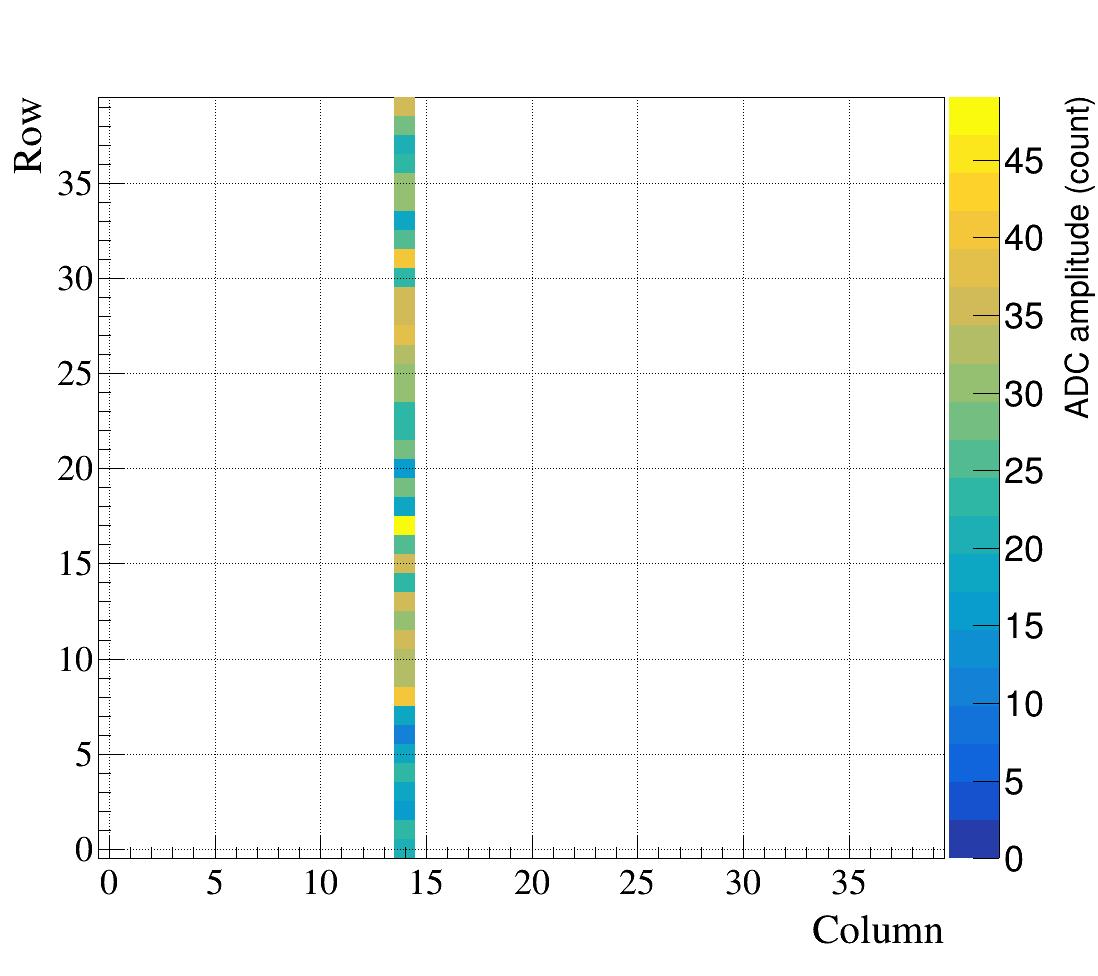}
\end{tabular}
\end{center}
\caption[example] 
{ \label{fig:cosmic_track} 
An example of a reconstructed cosmic-ray track in the ECAL. The color indicates the signal pulse amplitude in ADC counts.
}
\end{figure} 

\subsection{Calibration using cosmic rays}
\label{sec:calib_cosmic}
After verifying the performance of the ECAL modules using the light-monitoring system, cosmic-ray particles were used to establish initial high-voltage settings for each module. Relativistic cosmic-ray muons, which penetrate the detector modules, deposit a minimum amount of energy through ionization and are therefore referred to as minimum ionizing particles (MIPs). For a fixed path length, the energy deposited by a MIP is expected to be the same in all detector modules. Consequently, the HV applied to each module can be adjusted to equalize the signal pulse amplitudes produced by MIPs. 

\begin{figure} [htb]
\begin{center}
\begin{tabular}{c} 
  \includegraphics[width=10cm]{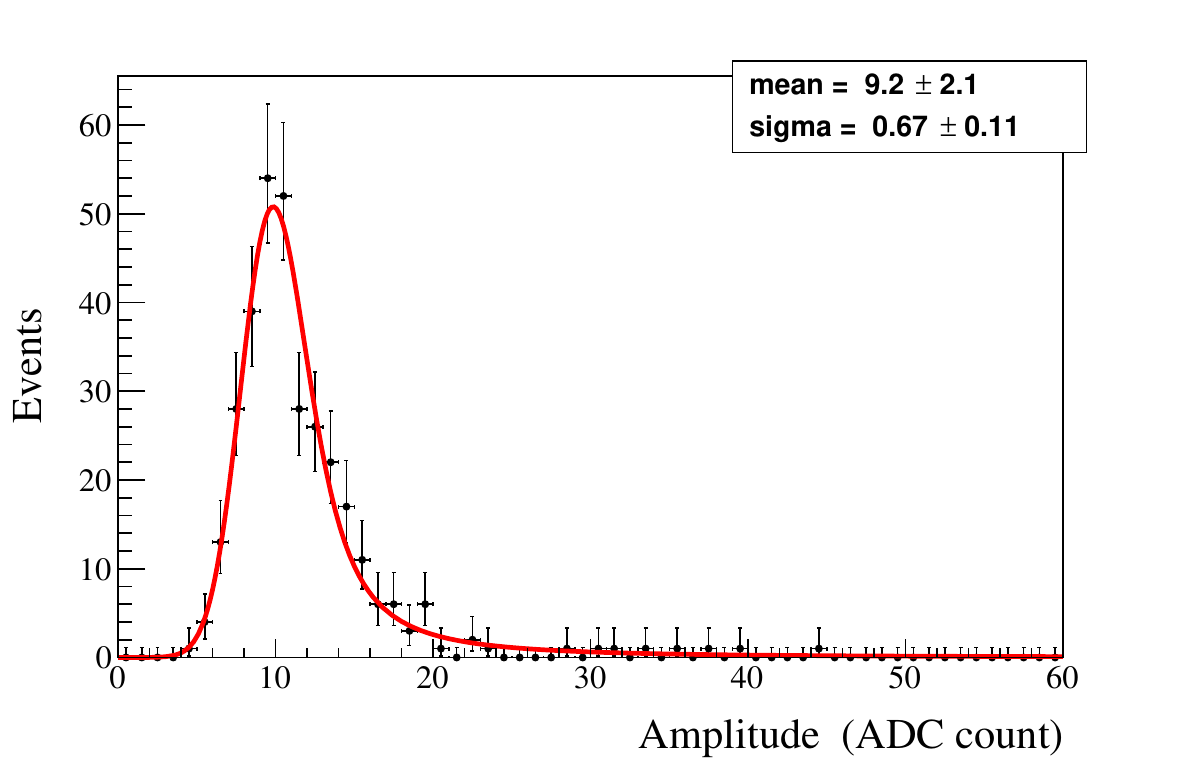}
\end{tabular}
\end{center}
\caption[example] 
{ \label{fig:cosmic_calib} 
Distribution of ADC amplitudes from cosmic-ray particles in an ECAL module. The solid line represents the fit.
}
\end{figure}

\begin{figure} [hbt]
\begin{center}
\begin{tabular}{c} 
  \includegraphics[width=10cm]{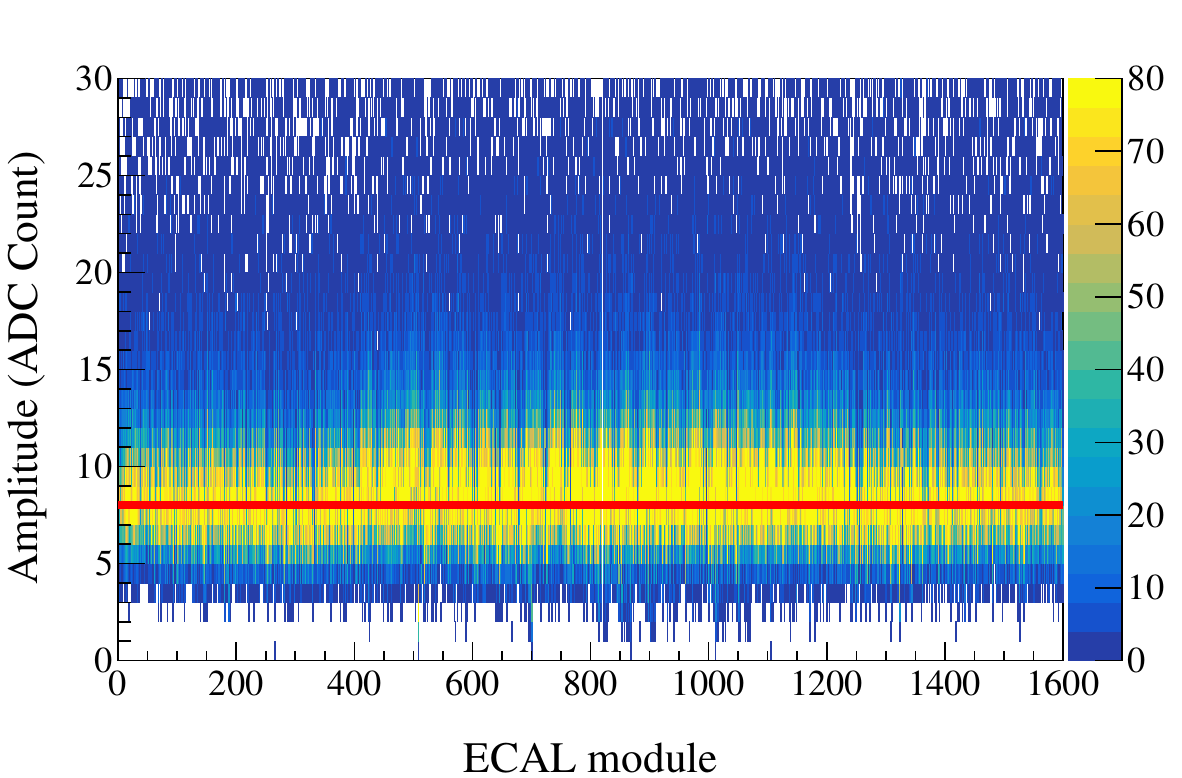}  
\end{tabular}
\end{center}
\caption[example] 
{ \label{fig:cosmic_calib1} 
Distribution of ADC amplitudes produced by cosmic rays in ECAL modules after high-voltage adjustment. The solid line indicates the target MIP response of 8  ADC counts.
}
\end{figure}

To detect cosmic-ray particles, four scintillating paddles were installed, with two placed above and two below the calorimeter, as shown in Fig.~\ref{fig:cosmic_setup}. The cosmic-ray trigger was formed using logic units located in a NIM crate. Signals from the upper and lower paddles were discriminated and then ORed within each group. A coincidence between the ORed top and bottom signals was required to generate a cosmic-ray trigger. The trigger signal was sent to the Trigger Supervisor module in the trigger crate to initiate the readout of the ECAL modules. For calibration, nearly vertical tracks in the ECAL were selected. An example of a reconstructed cosmic-ray track in the ECAL is shown in Fig.~\ref{fig:cosmic_track}.  A typical distribution of ADC pulse amplitudes recorded in an ECAL module for cosmic-ray  muons is shown in Fig.~\ref{fig:cosmic_calib}. This distribution was fitted with a Landau function convolved with a Gaussian to determine the average amplitude corresponding to the MIP response. The HVs of the ECAL modules were then adjusted to equalize these MIP amplitudes and to set the average MIP response to 8 ADC counts. This value was chosen  based on the full range of the 12-bit flash ADC and the expected energy deposition of cosmic-ray muons predicted by Geant4 detector simulation~\cite{geant4}, ensuring that higher-energy electromagnetic showers would not saturate the ADC. The resulting ADC amplitudes produced by cosmic-ray muons for all ECAL modules after HV adjustment are shown in Fig.~\ref{fig:cosmic_calib1}. The ECAL energy scale and module-to-module calibration were then refined using photon beam data.


\subsection{Calibration using photon beam}
\label{sec:calib_beam}

The subsequent stage of ECAL calibration was performed using a photon beam delivered to experimental Hall~D. This calibration utilized reconstructed Compton-scattering events, in which an incident beam photon scatters from a target electron via the reaction $\gamma + e \to \gamma^\prime + e^\prime$. The calibration procedure involved adjusting the global energy scale of the detector by tuning the PMT high voltages so that the signal amplitude produced by 10~GeV photons corresponded to 3500 ADC counts. In addition, the calibration refined the equalization of the energy response among ECAL modules, which had been initially calibrated using cosmic-ray data.
\begin{figure} [hbt]
\begin{center}
\begin{tabular}{c} 
\includegraphics[width=10cm]{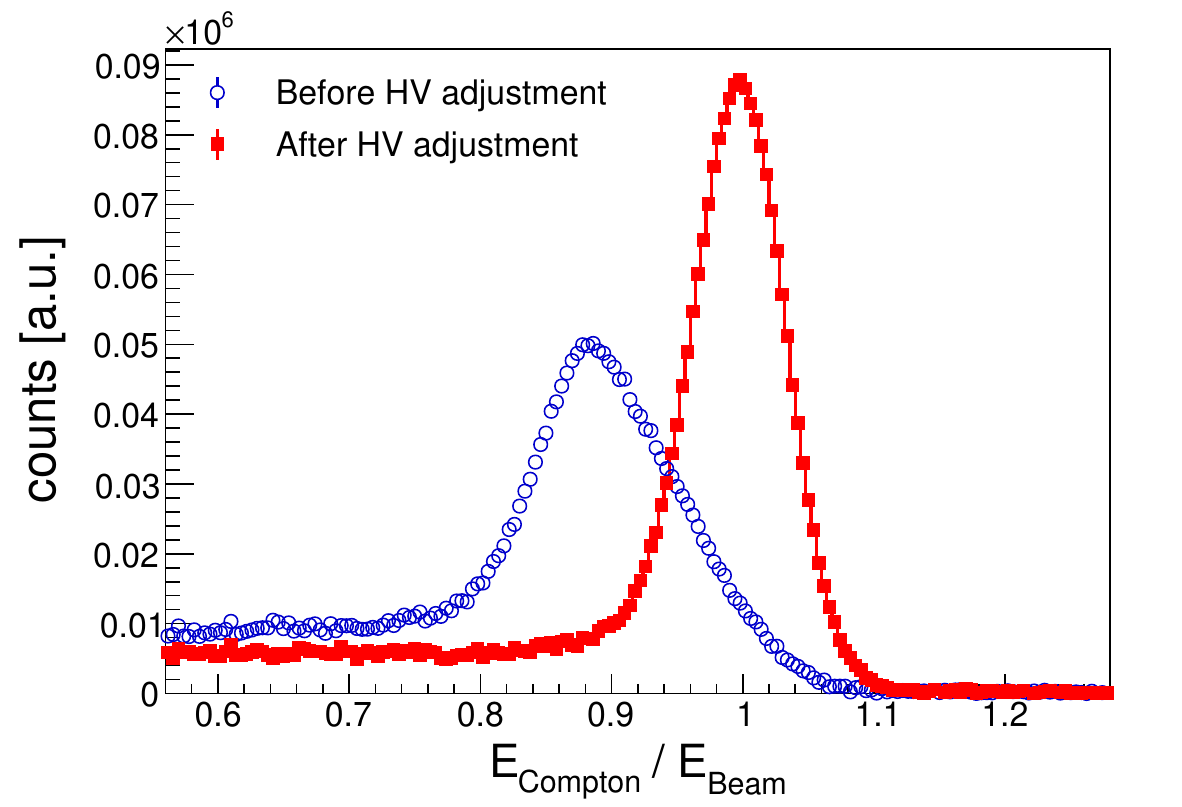}
\end{tabular}
\end{center}
\caption[example] 
{ \label{fig:gain_calib} 
Distribution of the ratio $R$ of reconstructed energy of Compton-scattering candidates to the beam energy.  Circular markers correspond to ECAL high-voltage settings obtained from cosmic-ray calibration, while square markers correspond to high voltages adjusted using Compton-scattering events.
}
\end{figure} 
The scattered photon and electron in the Compton-scattering process produce electromagnetic showers in the ECAL. Each shower deposits energy in multiple calorimeter modules, with the central module containing approximately $80\%$ of the total deposited energy. Candidate events were selected by requiring two electromagnetic showers in the calorimeter. The electron was distinguished from the photon by requiring an associated hit in the GlueX time-of-flight wall, a scintillator detector positioned upstream of the ECAL. Additional selection criteria were applied to suppress backgrounds from electromagnetic pair production and hadronic processes. These included timing requirements to ensure that the detected particles originated from the same beam bunch, as well as kinematic constraints of the Compton-scattering reaction.

\begin{figure} [htb]
\begin{center}
\begin{tabular}{c} 
\includegraphics[width=10cm]{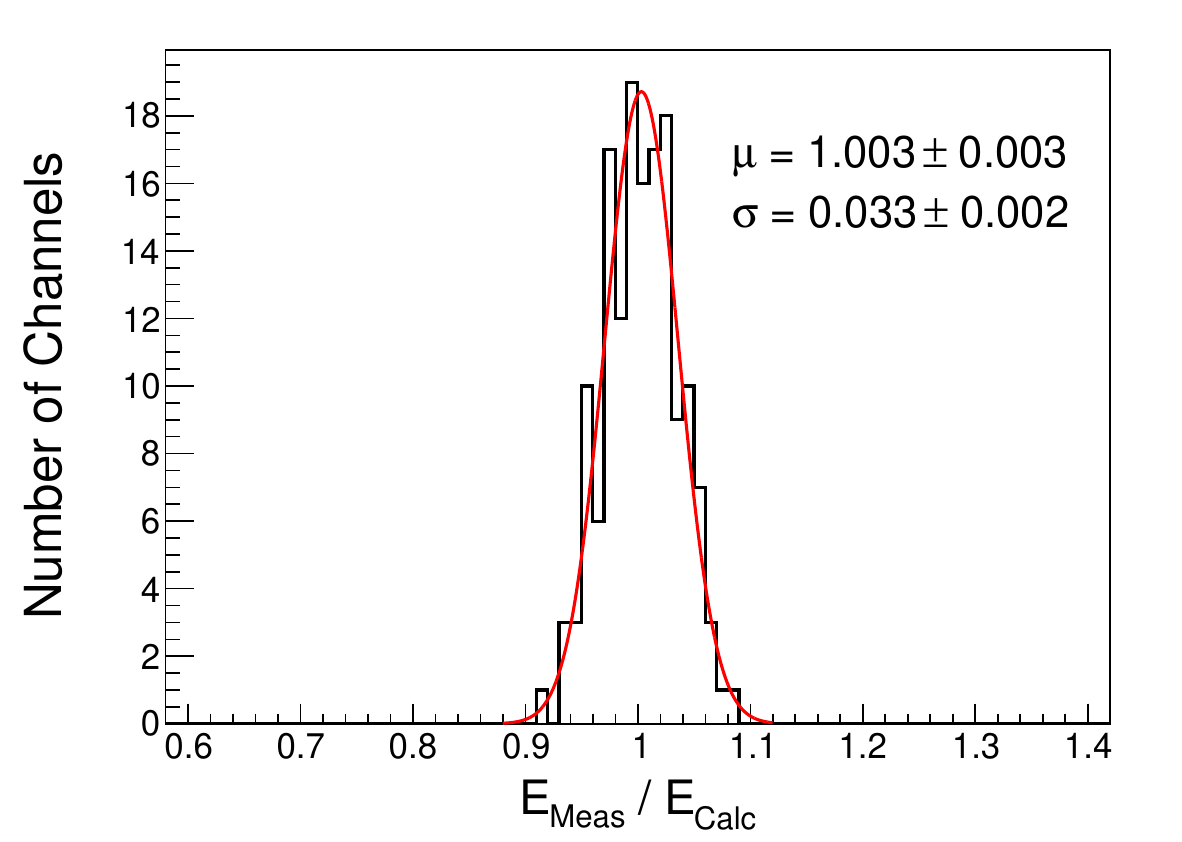}
\end{tabular}
\end{center}
\caption[example] 
{ \label{fig:gain_calib1} 
Distribution of the ratio of measured to calculated energy for Compton-scattering photons after ECAL high-voltage adjustment. The solid line shows a Gaussian fit to the distribution.
}
\end{figure} 

For the two-body Compton scattering process, the kinematics is fully determined by energy and momentum conservation. The energy of the scattered photon, $E_{\gamma^\prime}$, depends on the energy of the incident photon, $E_{\gamma}$, and the scattering angle $\theta$ according to
\begin{equation}
E_{\gamma^\prime} = \frac{E_\gamma}{1+\frac{E_\gamma}{m_e c^2}(1-\cos \theta )},
\label{eq:en}
\end{equation}
where $m_ec^2$ is the electron rest energy (0.511~MeV).
For each selected event, the expected energy of the scattered photon was calculated using Eq.~\ref{eq:en} from the measured scattering angle and compared with the shower energy measured in the calorimeter. The average ratio of the measured to calculated energy, 
\begin{equation}
R=E_{\rm Meas}/E_{\rm Calc},
\label{eq:en1}
\end{equation}
was determined for the central cell of each electromagnetic shower and used as a gain correction factor, such that the calibrated value of $R$ was unity. Gain corrections were applied only to the central cells of the showers, while the surrounding cells were left unchanged and used primarily for shower reconstruction. This procedure was repeated iteratively to achieve a uniform energy calibration across the detector.

The resulting gain correction factors were then used to adjust the PMT high voltages of the ECAL modules, accounting for variations in photomultiplier tube gain, light collection efficiency, and electronic response. Fig.~\ref{fig:gain_calib} shows the effect of the Compton-scattering calibration on the ECAL response, illustrating the distribution of the ratio of the reconstructed energy of Compton-scattering candidates to the incident beam energy. Circular markers correspond to  ECAL high-voltages settings obtained from the cosmic-ray calibration, while square markers correspond to high voltages adjusted using Compton-scattering events. 
Fig.~\ref{fig:gain_calib1} presents the distribution of gain factors, $R$, after the Compton calibration, along with a Gaussian fit. The relative width of the gain distribution is approximately 3\%. 

\begin{figure} [t]
\begin{center}
\begin{tabular}{c} 
  \includegraphics[height=7cm]{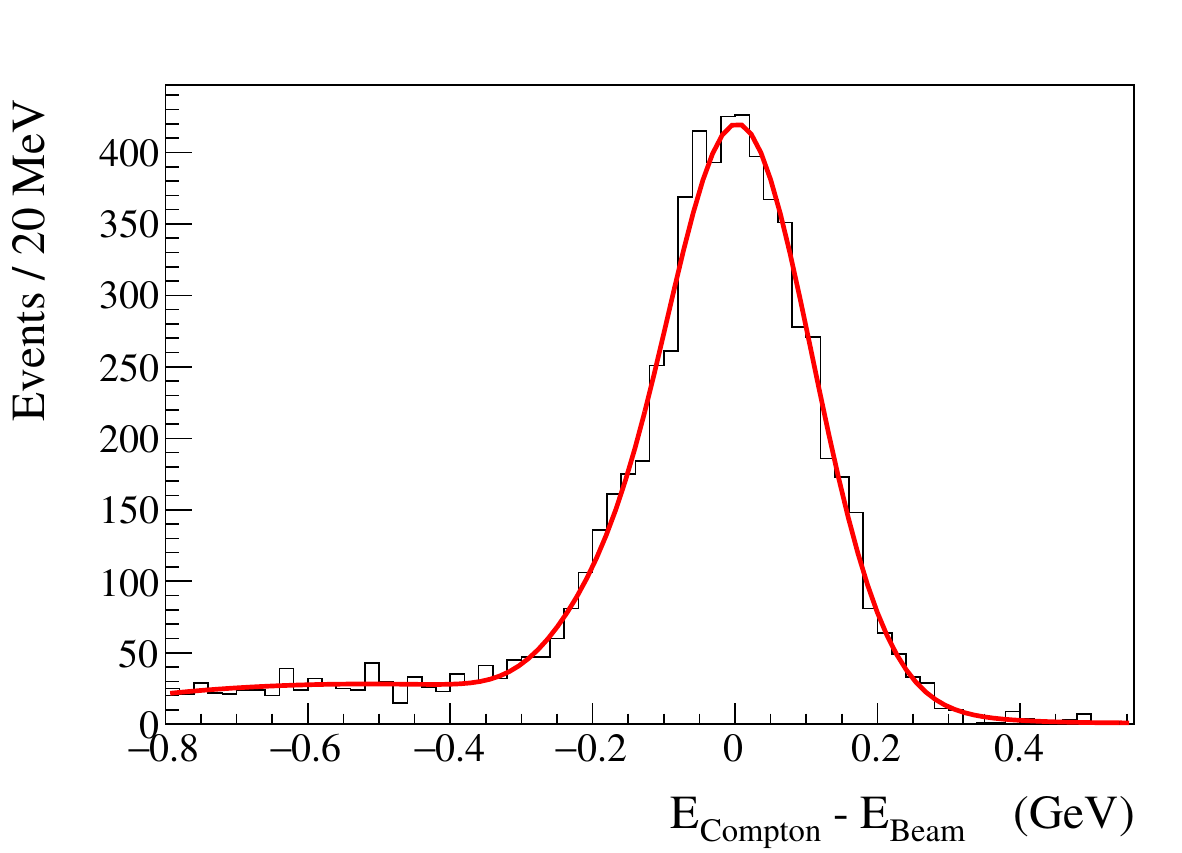}
\end{tabular}
\end{center}
\caption[] 
{ \label{fig:en_res} 
Distribution of the energy difference between reconstructed Compton-scattering candidates and the incident photon beam. The solid line shows a fit to the function described in the text. }
\end{figure} 

The difference between the reconstructed energy of Compton-scattering events and the beam energy for photon energies in the range between 6~GeV and 7~GeV is shown in Fig.~\ref{fig:en_res}. The distribution was fitted with the sum of two Gaussian function and a third-order polynomial, with the fit results superimposed on the data. The relative energy resolution of the reconstructed Compton events is approximately 1.7\%. 

Due to the relatively small scattering angles of the Compton photons, the calibration coverage is limited to approximately six ECAL layers surrounding the beam pipe. The final refinement of the ECAL calibration was performed using photons from decays of $\pi^0$ mesons. The calibration procedure based on the $\pi^0$ decays will be described in a forthcoming publication.

\section{Summary}
A large-scale calorimeter consisting of 1596 PbWO$_4$ scintillating crystals was fabricated, installed, and commissioned at Jefferson Lab for photon detection over a wide energy range at high event rates of up to 0.5 MHz per module. The detector was equipped with an LED-based light monitoring system and integrated into the data acquisition and trigger systems. Module optical responses were equalized using LED-based monitoring system, cosmic-ray muons, and Compton scattering events, and calibrated to the dynamic range of the detected photons. A calibration precision of 3\% was achieved, with stable performance observed during commissioning and subsequent data taking.

\section*{Acknowledgments}

This material is based upon work supported by the U.S. Department of Energy, Office of Science, Office of Nuclear Physics under contract DE-AC05-06OR23177, as well as NSF grants PHY-1812396, PHY-2111181, and PHY-2412800. We thank the Jefferson Lab Physics Division and members of the participating universities for their valuable assistance with ECAL fabrication and detector installation in the experimental hall. 


\bibliographystyle{plain}

\bibliography{ecal_article}

\end{spacing}
\end{document}